\documentclass[aps,pra,showpacs,floatfix]{revtex4}
\usepackage{amsfonts}
\usepackage{amsmath}
\usepackage{amssymb}
\usepackage{graphicx}

\begin{document}

\title{Anharmonic effects on a phonon number measurement of a quantum
mesoscopic mechanical oscillator}
\author{D. H. Santamore}
\affiliation{ITAMP, Harvard-Smithsonian Center for Astrophysics, Cambridge, MA 02138 USA}
\affiliation{Department of Physics, Harvard University, Cambridge, MA 02138 USA}
\affiliation{Department of Physics, California Institute of Technology, Pasadena, CA
91125 USA}
\author{Hsi-Sheng Goan}
\affiliation{Centre for Quantum Computer Technology, University of New South Wales,
Sydney, NSW 2052 Australia}
\author{G. J. Milburn}
\affiliation{Centre for Quantum Computer Technology, University of Queensland, St. Lucia,
QLD 4072 Australia}
\author{M. L. Roukes}
\affiliation{Condensed Matter Physics, California Institute of Technology, Pasadena, CA
91125 USA}

\begin{abstract}
We generalize a proposal for detecting single phonon transitions in a single
nanoelectromechanical system (NEMS) to include the intrinsic anharmonicity
of each mechanical oscillator. In this scheme two NEMS oscillators are
coupled via a term quadratic in the amplitude of oscillation for each
oscillator. One NEMS oscillator is driven and strongly damped and becomes a
transducer for phonon number in the other measured oscillator. We derive the
conditions for this measurement scheme to be quantum limited and find a
condition on the size of the anharmonicity. We also derive the relation
between the phase diffusion back-action noise due to number measurement and
the localization time for the measured system to enter a phonon number
eigenstate. We relate both these time scales to the strength of the measured
signal, which is an induced current proportional to the position of the
readout oscillator.
\end{abstract}

\date{\today}
\maketitle

\section{Introduction}

With device fabrication in the submicron or nanometer regime, it is possible
to fabricate mechanical oscillators with very high fundamental frequencies
and high mechanical quality factors. In the regime when the individual
mechanical quanta are of the order of or greater than the thermal energy,
quantum effects become important. Recently, a high-frequency mechanical
resonator beam that operates at GHz frequencies has been reported \cite%
{HZMR03}. Unlike quantum optical systems where extremely high frequency
oscillators, vacuum environments, zero temperature, and well-isolated
systems are the usual setup, solid state systems normally exist at finite
temperatures and interact with their surroundings. For a resonator operating
at the fundamental frequency of GHz and at a temperature of 100mK, on
average only 20 vibrational quanta are present in the fundamental mode. An
interesting question is: can we observe quantum jumps, i.e., discrete (Fock
or number state) transitions in such a true mechanical oscillator in a
mesoscopic solid system \cite{R00}, as the mechanical oscillator exchanges
quanta with the outside world or environment? In order to observe quantum
jumps, one needs to design a scheme to measure the phonon number of the
oscillator so that the oscillator will stay in a certain phonon number state
long enough before it jumps to another phonon number state due to the
inevitable interaction with its environment, usually through linear coupling
to the oscillator position.

To achieve a quantum mechanical phonon number measurement of a mechanical
oscillator, conventional measurement methods, such as the direct
displacement measurement, \cite{BG03} cannot be simply applied since the
observable (i.e., the number of phonons in the oscillator) does not commute
with, for example, the position or displacement operator. Thus, naively
attaching a readout transducer to the mechanical oscillator results in
inaccurate subsequent measurements due to back action. One thus must make
sure that the transducer that couples to the mechanical resonator measures
only the mean-square position, without coupling linearly to the resonator's
position itself \cite{R00}.

Some preliminary experiments in this direction have been conducted \footnote{%
For example, Ref.~\cite{HZMR03} provides high resonant frequency mechanical
oscillators. At the moment of this writing, the anharmonic coupling device
is being developed \cite{SP:Y}}. They use a second, driven mechanical
oscillator (oscillator $1$ in Fig.~\ref{model}) as the transducer to measure
the mean-square position of the system oscillator (oscillator $0$ in Fig.~%
\ref{model}). Hereafter, we use the notations of the \textquotedblleft
system oscillator\textquotedblright\ and \textquotedblleft ancilla
oscillator\textquotedblright\ in the text, but keep $0$\ and $1$\ as
subscripts in the mathematical notations. The basic idea is that the
non-linear, quadratic-in-position coupling between the two oscillators
shifts the resonance frequency of the ancilla oscillator by an amount
proportional to the phonon number or energy excitation of the system
oscillator. This frequency shift may be detected as a phase shift of the
oscillations of the ancilla oscillator with respect to the driving, when
driven at a fixed frequency near resonance. Also, the ancilla oscillator
needs to have sufficient sensitivity to resolve an individual quantum jump.

In the analysis of this measurement scheme presented by Santamore, Doherty,
and Cross \cite{SDC03}, self-anharmonic terms $x_{i}^{4}$, in the two
mechanical oscillators were neglected due to the smallness of the coupling
coefficients compared to their harmonic oscillation frequencies, where $%
x_{i} $, $i=0,1$ is the displacement of the oscillators position from
equilibrium. Since the self-anharmonic terms are of the same order as the
nonlinear coupling term $x_{0}^{2}x_{1}^{2}$, it is important to include
those terms and analyze the effects on the proposed measurement scheme.

In this paper, we extend the work of Ref.\ \cite{SDC03} and investigate the
effects of self-anharmonic terms\ on a phonon number measurement. Due to the
higher order self-anharmonic terms, the adiabatic elimination method used in
Ref.\ \cite{SDC03} may not be straightforwardly applied even with the
assumption of a heavily damped ancilla oscillator due to measurement. Here
we take a slightly different approach. As the ancilla is assumed to be
heavily damped, it will relax very rapidly to its steady state within a
timescale on the order of typical response time of the system oscillator,
and will appear to the system oscillator effectively as a \textquotedblleft
bath\textquotedblright . To see the consequences of a rapidly decaying
ancilla oscillator on the dynamics of the system oscillator, we use the
quantum open systems approach to find the master equation for the reduced
density matrix of the system oscillator. In obtaining the master equation,
the correlation functions of the \textquotedblleft effective
bath\textquotedblright\ (or the ancilla oscillator) are calculated using the
generalized P-representation approach \cite{DG80}. The generalized
P-representation approach has the advantage of removing some of the
unnecessary restrictions imposed in Ref.\ \cite{SDC03}.

We find that in the presence of self-anharmonic term, $x_1^{4}$, of the
ancilla oscillator, the effect of increasing driving strength and
self-nonlinearity tends to shift the resonance frequency, increase the peak
value and decrease the width of the response of the peak of $(\Gamma /\Gamma
_{0})$ (see Fig.~\ref{QJ_Gamma}). The quantity $(\Gamma /\Gamma _{0})$ is
the ratio of the back-action diffusion coefficient (or decoherence rate) $%
\Gamma $ [see Eq.~(\ref{eq:rateG})] and its value $\Gamma _{0}$ at zero
self-anharmonicity and zero detuning. If the damping of the ancilla
oscillator is much larger than the effect of the self-anharmonic term, the
overall effect of self-anharmonic term on the phonon number measurement is
small. Finally, we show that the induced electromotive readout current \cite%
{YGPB94} from the ancilla oscillator provides information on the phonon
number of the system, even in the presence of higher order anharmonic terms,
and we obtain the relation between the current and the measured system
observable.

In the next section, we discuss briefly the measurement scheme and
Hamiltonian, and obtain the master equation for the model described above
while keeping higher order self-anharmonic terms. It turns out that the
master equation we obtain requires two-time correlations of the ancilla
oscillator operators. Section \ref{Sec_correlation} deals with this issue.
We find one-time and two-time correlation functions of the ancilla. In Sec.\ %
\ref{Sec_analysis}, we examine the effect of the self-anharmonic terms on
the dynamics of the system oscillator from the master equation of its
reduced density matrix. In Sec.\ \ref{Sec_current}, we obtain the dependence
of the measurement current on the measured system oscillator observable, the
phonon number.

\section{Hamiltonian and the master equation}

\subsection{Proposed scheme}

\begin{figure}[tbh]
\begin{center}
\includegraphics[width=4.5in]{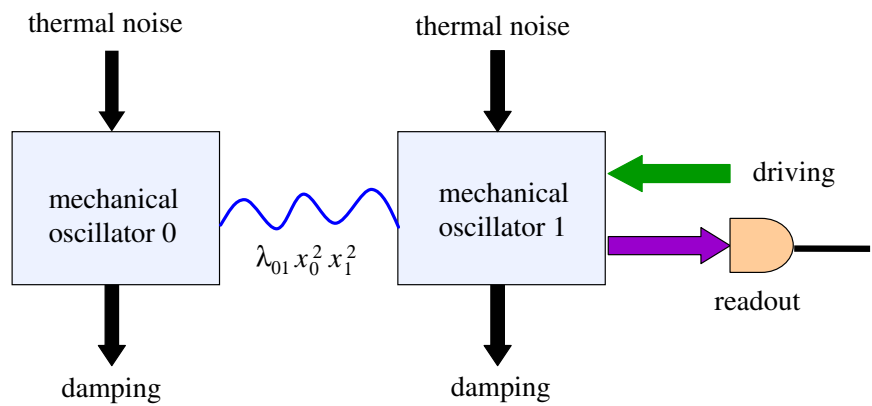}
\end{center}
\caption{Schematic of phonon number measurement for a coupled mechanical
oscillator. Oscillators $0$ and $1$ are anharmonically coupled with coupling
strength $\protect\lambda _{01}$. Both oscillators are subjected to thermal
noise injection and dissipation. The oscillator $1$ is driven and a readout
apparatus is attached to it.}
\label{model}
\end{figure}

Our model consists of two mesoscopic scale mechanical bridges with
rectangular cross section. One serves as a system oscillator (oscillator 0
in Fig.~\ref{model}) to be measured. The other is used as an ancilla
oscillator (oscillator 1 in Fig.~\ref{model}), and is part of the measuring
apparatus. Details of the scheme have been already discussed in Ref.~\cite%
{SDC03}. A schematic illustration is reproduced in Fig.~\ref{model}. These
mesoscopic-size elastic bridges or beams with rectangular cross-section are
connected by a device that transmits only one of the flexing modes of the
system oscillator to the ancilla oscillator. As a result, these two
resonators are anharmonically and symmetrically coupled (for experimental
progress toward the scheme, see Ref.~\cite{HZMR03}). We label the measured
system oscillator with subscript $0$ and the ancilla oscillator with
subscript $1$, with corresponding resonant frequencies of the two flexing
modes labelled as $\omega _{0}$ and $\omega _{1}$, respectively. The ancilla
oscillator is driven at frequency $\omega _{d}$ with strength $\epsilon (t)$%
. A measuring apparatus is attached to the ancilla oscillator. The whole
structure is subjected to the thermal bath environment. The interaction of
the system oscillator with the thermal bath causes thermal dissipation and
excitation of the system oscillator, which results in random-in-time
transitions between phonon number eigenstates (i.e., quantum jumps). A
change in the energy of the system oscillator appears to the ancilla
oscillator as a shift of the resonant frequency via the anharmonic coupling.
This frequency shift may be detected as a phase shift of the oscillations of
the ancilla oscillator with respect to the driving, when driven at a fixed
frequency near resonance.

\subsection{Model Hamiltonian}

The free Hamiltonian for the two bridge oscillators $0$ and $1$ is
\begin{equation}
H_{\mathrm{free}}=\hbar \omega _{0}a^{\dagger }a+\hbar \omega _{1}b^{\dagger
}b,
\end{equation}%
where $a^{\dagger },a$ are creation and annihilation operators for
oscillator $0$, respectively, and similarly, $b^{\dagger },b$ for oscillator
$1$. The ancilla oscillator is driven at frequency $\omega _{d}$ with
driving strength $\epsilon $,
\begin{equation}
H_{\mathrm{drive}}=\hbar \epsilon \cos \left[ \omega _{d}t\right] \left(
b^{\dagger }+b\right) ,
\end{equation}%
In the interaction picture, the driving term becomes
\begin{equation}
H_{\mathrm{drive}}^{\mathrm{I}}=2\hbar \epsilon \left( b^{\dagger
}e^{+i\delta \omega }+be^{-i\delta \omega }\right) ,
\end{equation}%
where $\delta \omega $ is detuning between the ancilla resonant frequency
and the driving frequency, $\omega _{1}-\omega _{d}$.

The two oscillators are coupled anharmonically through the special coupling
device that controls and allows only one type of strain (the longitudinal
stretch) to pass to other oscillator. Beyond the linear elasticity theory,
the two flexing modes, which are perpendicular to each other, are coupled.
Expansion of the elastic energy with respect to the strain tensor is taken
up to second order. The next term, cubic in the elastic energy, gives
quadratic terms in the equation of motion \cite{LL,T}. Since the coupling of
the two modes of the two beams is symmetric, and since the two modes are not
coupled at the linear level, the first order in coupling is $%
x_{0}^{2}x_{1}^{2}$, where $x_{i}$\ is the displacement operator. So we
expand the anharmonic terms up to first order in coupling and obtain
\begin{align}
H_{anh}& =\hbar \left( \tilde{\lambda}_{0}x_{0}^{2}+\tilde{\lambda}%
_{00}x_{0}^{4}+\tilde{\lambda}_{1}x_{1}^{2}+\tilde{\lambda}%
_{11}x_{1}^{4}\right) ,  \label{eq:pure-anharmonic} \\
V_{anh}& =\hbar \tilde{\lambda}_{01}x_{0}^{2}x_{1}^{2},
\label{eq:int-anharmonic}
\end{align}%
where $\tilde{\lambda}_{ij}$ is the coupling coefficient. The high
frequencies of the resonators, i.e., $(\omega _{0}-\omega _{1})$ much larger
than $\tilde{\lambda}_{01}$ and their damping rates, allows us to use the
rotating wave approximation. Thus we write the anharmonic terms as
\begin{align}
H_{anh}& =\hbar \lambda _{00}\left( a^{\dagger }a\right) ^{2}+\hbar \lambda
_{11}\left( b^{\dagger }b\right) ^{2},  \label{vnaharmonic} \\
V_{anh}& =\hbar \lambda _{01}a^{\dagger }ab^{\dagger }b,
\end{align}%
where we have defined the standard raising and lowering operators for the
oscillators: $a=\sqrt{m_{0}\omega _{0}/2\hbar }\;x_{0}+i\sqrt{1/2\hbar
m_{0}\omega _{0}}\;p_{0}$, $a^{\dagger }$ is the Hermitian conjugate of $a$
and similarly for $b$ and $b^{\dagger }$ with the subscript $0$ replaced by $%
1$. We have also introduced new coefficients $\lambda $'s (without tildes)
which all have the same dimension of frequency.

The coupling term $\lambda _{01}a^{\dagger }ab^{\dagger }b$\ commutes with
the observable $a^{\dagger }a$, enabling a quantum non-demolition (QND)
measurement. The terms $\lambda _{0}a^{\dagger }a$ and $\lambda
_{1}b^{\dagger }b$ shift the resonance frequency by a constant amount, so we
have absorbed these quantities into $\omega _{0}$ and $\omega _{1}$. The
terms $\hbar \lambda _{00}(a^{\dagger }a)^{2}$ and $\hbar \lambda
_{11}(b^{\dagger }b)^{2}$ are analogous to Kerr non-linearities in nonlinear
optics. Since these terms commute with the measured observable $(a^{\dagger
}a)$, they will not change the system phonon number eigenstates; however,
the Kerr effect causes an intensity dependent phase shift. Unlike a coherent
state, in which this effect results in rotational shearing, a thermal state
will not be affected by phase shift, due to its rotational invariance.

As for detecting phonon number in the system oscillator, we adapt a
magnetomotive detection scheme suggested by Yurke et al \cite%
{YGPB94,CR96,CR99}. The voltage developed is proportional to $dx_{1}/dt$,
where $x_{1}$\ is the displacement of the beam from its equilibrium
position. The current induced by this voltage is monitored by phase lock-in
amplifier. An experimenter monitors the amplitude of the current and its
phase with respect to the driving current that is set to a frequency near
resonance. The details of the relation between the measured current and the
phonon number of the system oscillator are derived in Ref.\ (\cite{SDC03}).

There are two physically distinct environments in the model: the
thermo-mechanical environment of each oscillator and the electronic noise
environment of the electrical system that ultimately provides information on
the motion of the ancilla. The environments are modelled as thermal baths,
each consisting of an infinite number of harmonic oscillators. The couplings
between the oscillators and the thermal baths are considered as weak, linear
and Markovian; thus we use the rotating wave approximation. The Hamiltonian
of the baths and their coupling to the oscillators can then be written as:
\begin{eqnarray}
H_{bath} &=&\hbar \sum_{\mathrm{s}}\sum_{n}\omega _{\mathrm{s},i}B_{\mathrm{s%
},n}^{\dagger }B_{\mathrm{s},n}, \\
V_{bath} &=&\hbar \left( \Omega _{B0}^{\dagger }\,a+a^{\dagger }\Omega
_{B0}\right) +\hbar \left( \Omega _{B1}^{\dagger }\,b+b^{\dagger }\Omega
_{B1}\right) +\hbar \left( \Omega _{Bm}^{\dagger }\,b+b^{\dagger }\Omega
_{Bm}\right),
\end{eqnarray}%
where $s$ runs over three different baths: the thermal baths coupled to the
system oscillator ($B0$) and ancilla oscillator ($B1$), and the electronic
(measurement) bath coupled to ancilla oscillator ($Bm$). The operator
\begin{equation}
\Omega _{\mathrm{s}}=\sum_{\mathrm{n}}g_{\mathrm{s}}\left( \omega
_{n}\right) B_{s,n}  \label{eq:bathcoupling}
\end{equation}%
consists of bath operators, and the coupling to the bath modes is given by
the coefficients $g_{\mathrm{s}}\left( \omega _{n}\right) $.

\subsection{Master equation}

Using the standard technique for open quantum systems, we first obtain the
master equation for the joint density matrix of the two oscillators, $R$, by
tracing out the bath variables:
\begin{align}
\frac{dR}{dt}& =-i\omega _{0}\left[ a^{\dagger }a,R\right] -i\lambda _{00}%
\left[ \left( a^{\dagger }a\right) ^{2},R\right]  \notag \\
& -i\delta \omega \left[ b^{\dagger }b,R\right] -i\epsilon \left[ b^{\dagger
}+b,R\right] -i\lambda _{11}\left[ \left( b^{\dagger }b\right) ^{2},R\right]
-i\lambda _{01}\left[ a^{\dagger }ab^{\dagger }b,R\right]  \notag \\
& +\nu \left( N_{0}+1\right) \mathcal{D}\left[ a\right] R+\nu N_{0}\mathcal{D%
}\left[ a^{\dagger }\right] R+\kappa \left( N_{1}+1\right) \mathcal{D}\left[
b\right] R+\kappa N_{1}\mathcal{D}\left[ b^{\dagger }\right] R,
\label{D-master-eqn}
\end{align}%
where
\begin{align}
\mathcal{D}\left[ O\right] R& =2ORO^{\dagger }-(O^{\dagger }OR+RO^{\dagger
}O), \\
\mathcal{D}\left[ O^{\dagger }\right] R& =2O^{\dagger }RO-(OO^{\dagger
}R+ROO^{\dagger }).
\end{align}%
are defined for arbitrary operators $O$ and $R$. The damping rate of the
system oscillator $\nu $ is given by
\begin{equation}
\nu \equiv \pi \varrho _{\mathrm{B0}}\left( \omega _{0}\right) \left\vert g_{%
\mathrm{B0}}\left( \omega _{0}\right) \right\vert ^{2}.  \label{eq:defnu}
\end{equation}%
It is related to the quality factor $Q_{0}$ of the system oscillator by $\nu
=\omega _{0}/2Q_{0}$. We have combined the damping rates $\mu $ and $\eta $,
due respectively to thermal bath and measurement on ancilla oscillator, into
$\kappa =\mu +\eta $, where
\begin{align}
\mu & \equiv \pi \varrho _{\mathrm{Bm}}\left( \omega _{1}\right) \left\vert
g_{\mathrm{Bm}}\left( \omega _{1}\right) \right\vert ^{2},  \label{eq:defmu}
\\
\eta & \equiv \pi \varrho _{\mathrm{B1}}\left( \omega _{1}\right) \left\vert
g_{\mathrm{B1}}\left( \omega _{1}\right) \right\vert ^{2}.  \label{eq:defeta}
\end{align}%
Here $\varrho _{\mathrm{s}}\left( \omega \right) $ is the density of states
of bath $s$ at frequency $\omega $. The $N_{\mathrm{i}}$ are the
Bose-Einstein factors:
\begin{equation}
N_{0}=\frac{1}{e^{\hbar \beta _{0}\omega _{0}}-1},
\end{equation}%
and $N_{1}=\left( \eta N_{\bar{1}}+\mu N_{\mathrm{m}}\right) /\kappa $,
where
\begin{equation}
N_{\bar{1}}=\frac{1}{e^{\hbar \beta _{1}\omega _{1}}-1},\ N_{\mathrm{m}}=%
\frac{1}{e^{\hbar \beta _{\mathrm{m}}\omega _{1}}-1},
\end{equation}%
with $\beta _{\mathrm{i}}=\left( k_{\mathrm{B}}T_{s}\right) ^{-1}$ and $%
T_{s} $ the temperature of bath $s$. In Eq.\ (\ref{D-master-eqn}), the first
and the second lines are the free Hamiltonian and non-linear Kerr effect
terms for system and ancilla oscillators, respectively.
%For convenience, we use subscript $0$ for system and $1$ for ancilla
%in the mathematical notations and equations.
The third line in Eq.\ (\ref{D-master-eqn}) is associated with the
anharmonic coupling, and the last two lines are consequences of the
interactions with thermal baths.

\section{Effect of heavily damped ancilla oscillator}

To proceed further towards a master equation for the reduced density matrix
for the system oscillator alone, the ancilla oscillator is assumed to be
heavily damped due to measurements, i.e., $\kappa \gg \lambda _{ij},\nu $.
In this case, the ancilla oscillator will relax very rapidly to its steady
state and appear to the system oscillator as a \textquotedblleft
bath\textquotedblright . In fact, if $\lambda _{11}\ll \omega _{1}$ and $%
\lambda _{01}\ll \kappa $, the ancilla oscillator in Eq.~(\ref{D-master-eqn}%
) will remain near a thermal steady state with average number $N_{1}$.
However, we will relax the condition $\lambda _{11}\ll \omega _{1}$ and
treat the interaction $\lambda _{01}$ term pertubatively.

To see the consequences of the rapid decay of the ancilla oscillator on the
dynamics of the system oscillator, we use perturbation theory and expand the
interaction Hamiltonian $H_{I}\left( t\right) =\lambda _{01}a^{\dagger
}ab^{\dagger }b$ up to second order, and trace out the ancilla oscillator
variables. This implies that we need to calculate the relevant steady state
averages and correlation functions for the ancilla oscillator in the
presence of the anharmonic term $\lambda _{11}\left( b^{\dagger }b\right)
^{2}$.

In this case, the master equation for the reduced density matrix $\rho (t)$\
for the system oscillator alone can be written as
\begin{align}
\frac{d\rho (t)}{dt}& =-i\omega _{0}\left[ a^{\dagger }a,\rho (t)\right]
-i\lambda _{00}\left[ \left( a^{\dagger }a\right) ^{2},\rho (t)\right] +\nu
\left( N_{0}+1\right) \mathcal{D}\left[ a\right] \rho (t)+\nu N_{0}\mathcal{D%
}\left[ a^{\dagger }\right] \rho (t)  \notag \\
& -i\text{Tr}_{1}\left[ H_{I}\left( t\right) ,R_{\mathrm{eff}}(t)\right]
-\int_{0}^{t}\text{Tr}_{1}\left[ H_{I}\left( t\right) ,\left[ H_{I}\left(
t^{\prime }\right) ,R_{\mathrm{eff}}(t)\right] \right] dt^{\prime },
\label{MasterEqint}
\end{align}%
where $R_{\mathrm{eff}}(t)\approx \rho (t)\rho _{1}(t)$ is the effective
joint density matrix of the two oscillators under the approximation that the
ancilla oscillator is heavily damped, and $\rho _{1}(\infty )$ is the steady
state density matrix operator for the ancilla oscillator. Explicitly, the
second term of the last line of Eq.\ (\ref{MasterEqint}) can be written as
\begin{align}
& \int_{0}^{t}\text{Tr}_{1}\left[ H_{I}\left( t\right) ,\left[ H_{I}\left(
t^{\prime }\right) ,R_{\mathrm{eff}}(t)\right] \right] dt^{\prime }  \notag
\\
& =-\left( \lambda _{01}\right) ^{2}\int_{0}^{t}a^{\dagger }a\left( t\right)
a^{\dagger }a\left( t^{\prime }\right) \rho (t)\left\langle b^{\dagger
}b\left( t\right) b^{\dagger }b\left( t^{\prime }\right) \right\rangle
dt^{\prime }+\left( \lambda _{01}\right) ^{2}\int_{0}^{t}a^{\dagger }a\left(
t\right) \rho (t)a^{\dagger }a\left( t^{\prime }\right) \left\langle
b^{\dagger }b\left( t^{\prime }\right) b^{\dagger }b\left( t\right)
\right\rangle dt^{\prime }  \notag \\
& +\left( \lambda _{01}\right) ^{2}\int_{0}^{t}a^{\dagger }a\left( t^{\prime
}\right) \rho (t)a^{\dagger }a\left( t\right) \left\langle b^{\dagger
}b\left( t\right) b^{\dagger }b\left( t^{\prime }\right) \right\rangle
dt^{\prime }-\left( \lambda _{01}\right) ^{2}\int_{0}^{t}\rho (t)a^{\dagger
}a\left( t^{\prime }\right) a^{\dagger }a\left( t\right) \left\langle
b^{\dagger }b\left( t^{\prime }\right) b^{\dagger }b\left( t\right)
\right\rangle dt^{\prime }.  \label{2ndOrder}
\end{align}%
The exact correlation functions of the ancilla oscillator are not easy to
evaluate because of the presence of the anharmonicity, the driving, and the
decay terms. However, one can make an expansion of the state of the ancilla
oscillator around its steady state and linearize the fluctuations, assuming
them to be small \cite{WM,DW80}.

Define the steady-state mean field amplitudes as $\left\langle
b\right\rangle _{\infty }=\beta _{0}$. The operator $b$ can be written in
terms of small fluctuations about the steady state mean value as
\begin{equation}
b\left( t\right) =\beta _{0}+b_{1}\left( t\right) .
\end{equation}
Then, keeping terms up to quadratic order in $b_{1},b_{1}^{\dagger },$ the
interaction Hamiltonian $H_{I}=\lambda _{01}a^{\dagger }ab^{\dagger }b$
becomes
\begin{equation}
H_{I}=\lambda _{01}a^{\dagger }a\left[ \left\vert \beta _{0}\right\vert
^{2}+\beta _{0}^{\ast }b_{1}\left( t\right) +\beta _{0}b_{1}^{\dagger
}\left( t\right) +b_{1}^{\dagger }\left( t\right) b_{1}\left( t\right) %
\right] .  \label{eq:Hintanharmonic}
\end{equation}
The first term in Eq.\ (\ref{eq:Hintanharmonic}) contributes to a shift in
the resonant frequency of the system oscillator by a constant amount and can
be combined with the free Hamiltonian. Inserting this expression back into
the first term of the last line of Eq.~(\ref{MasterEqint}) gives the first
order expansion term
\begin{equation}
-i\lambda _{01}\text{Tr}_{1}\left[ a^{\dagger }a\left( t\right) b^{\dagger
}b\left( t\right) ,R_{\mathrm{eff}}(t)\right] =-i\lambda _{01}\left[
a^{\dagger }a\left( t\right) ,\rho \right] \left\langle b_{1}^{\dagger
}b_{1}(t)\right\rangle ,  \label{eq:1st order}
\end{equation}
where we have used the fact that averages of fluctuation fields vanish,
i.e.,
\begin{equation}
\left\langle b_{1}\right\rangle =\left\langle b_{1}^{\dagger }\right\rangle
=0.
\end{equation}

Now we turn our attention to the second order term, Eq.\ (\ref{2ndOrder}).
Note that since $\kappa \gg \nu $, the phonon number $a^{\dagger }a(t)$ of
the system oscillator changes with time on a time scale much larger than $%
b^{\dagger }b(t)$ of the ancilla oscillator. So we can approximate $%
a^{\dagger }a(t^{\prime })\simeq a^{\dagger }a(t)$ in Eq.\ (\ref{2ndOrder})
and pull the system oscillator terms outside of the integral. Then Eq.\ (\ref%
{2ndOrder}) becomes
\begin{align}
& \int_{0}^{t}\text{Tr}_{1}\left[ H_{I}\left( t\right) ,\left[ H_{I}\left(
t^{\prime }\right) ,R_{\mathrm{eff}}(t)\right] \right] dt^{\prime }  \notag
\\
& \approx \left( \lambda _{01}\right) ^{2}\left\{ a^{\dagger }a\left(
t\right) \rho (t)a^{\dagger }a\left( t\right) -\left( a^{\dagger }a\left(
t\right) \right) ^{2}\rho (t)\right\} \int_{0}^{t}\left\langle B\left(
t,t^{\prime }\right) \right\rangle dt^{\prime }  \notag \\
& +\left( \lambda _{01}\right) ^{2}\left\{ a^{\dagger }a\left( t\right) \rho
(t)a^{\dagger }a\left( t\right) -\rho (t)\left( a^{\dagger }a\left( t\right)
\right) ^{2}\right\} \int_{0}^{t}\left\langle B\left( t^{\prime },t\right)
\right\rangle dt^{\prime }  \label{2nd order}
\end{align}%
where
\begin{align}
\left\langle B\left( t,t^{\prime }\right) \right\rangle & =\left( \beta
_{0}^{\ast }\right) ^{2}\left\langle b_{1}\left( t\right) b_{1}\left(
t^{\prime }\right) \right\rangle +\left\vert \beta _{0}\right\vert
^{2}\left\langle b_{1}\left( t\right) b_{1}^{\dagger }\left( t^{\prime
}\right) \right\rangle +\left\vert \beta _{0}\right\vert ^{2}\left\langle
b_{1}^{\dagger }\left( t\right) b_{1}\left( t^{\prime }\right) \right\rangle
+\left( \beta _{0}\right) ^{2}\left\langle b_{1}^{\dagger }\left( t\right)
b_{1}^{\dagger }\left( t^{\prime }\right) \right\rangle \\
\left\langle B\left( t^{\prime },t\right) \right\rangle & =\left( \beta
_{0}^{\ast }\right) ^{2}\left\langle b_{1}\left( t^{\prime }\right)
b_{1}\left( t\right) \right\rangle +\left\vert \beta _{0}\right\vert
^{2}\left\langle b_{1}\left( t^{\prime }\right) b_{1}^{\dagger }\left(
t\right) \right\rangle +\left\vert \beta _{0}\right\vert ^{2}\left\langle
b_{1}^{\dagger }\left( t^{\prime }\right) b_{1}\left( t\right) \right\rangle
+\left( \beta _{0}\right) ^{2}\left\langle b_{1}^{\dagger }\left( t^{\prime
}\right) b_{1}^{\dagger }\left( t\right) \right\rangle
\end{align}%
and higher order fluctuation terms than $b_{1}^{2}$ are ignored. The
linearization transforms the second-order correlation functions of the
ancilla operators, $\langle b^{\dagger }b(t)b^{\dagger }b(t^{\prime
})\rangle $ and $\left\langle b^{\dagger }b\left( t^{\prime }\right)
b^{\dagger }b\left( t\right) \right\rangle $, into first order correlation
functions of fluctuation fields: $\langle b_{1}^{\dagger }\left( t\right)
b_{1}\left( t^{\prime }\right) \rangle $, $\langle b_{1}^{\dagger }\left(
t\right) b_{1}^{\dagger }\left( t^{\prime }\right) \rangle $ $\langle
b_{1}^{\dagger }\left( t\right) b_{1}\left( t^{\prime }\right) \rangle $,
and $\langle b_{1}^{\dagger }\left( t\right) b_{1}^{\dagger }\left(
t^{\prime }\right) \rangle $.

\section{One-time and two-time correlation functions of ancilla\label%
{Sec_correlation}}

In this section we calculate the one-time and two-time correlation functions
of the ancilla oscillator. For this purpose, first we need to calculate the
one-time correlation functions of a single driven anharmonic oscillator. We
will follow the method of Drummond and Walls \cite{DW80}, who obtained
one-time correlation functions. Then we extend their method to calculate
two-time correlation functions.

The master equation for the driven, anharmonic ancilla oscillator
interacting with the thermal bath is given by
\begin{align}
\frac{d{\rho _{1}(t)}}{dt}& =-i\delta \omega \left[ b^{\dagger }b,\rho
_{1}(t)\right] -i\epsilon \left[ b^{\dagger }+b,\rho _{1}(t)\right]
-i\lambda _{11}\left[ \left( b^{\dagger }b\right) ^{2},\rho _{1}(t)\right]
\notag \\
& +\kappa \left( N_{1}+1\right) \mathcal{D}\left[ b\right] \rho
_{1}(t)+\kappa N_{1}\mathcal{D}\left[ b^{\dagger }\right] \rho _{1}(t).
\label{master-anharmonic}
\end{align}
where $\rho _{1}$ is the density matrix of the ancilla oscillator and $%
\delta \omega=\omega _{1}-\omega _{d}$ is the detuning, with $\omega _{d}$
the driving frequency. The exact steady-state one-time correlation functions
for a system with master equation Eq.\ (\ref{master-anharmonic}) at \textit{%
zero temperature} were given in Refs.~\cite{WM,DW80}, in a discussion of
optical bistability of a coherently driven dispersive cavity with a cubic
nonlinearity in the polarizability of the internal medium. At finite
temperature, no exact solution has been found.

Our first objective is to derive a stochastic differential equation from the
quantum master equation. Representing a density matrix in a coherent state
basis is useful in systems described by Bose operators $b^{\dagger},b$. Due
to the presence of the non-linear, self-anharmonic term, we will use the
generalized P-representation introduced by Drummond and Gardiner \cite{DG80}
to preserve the positivity of the Hermitian density operator.

Using the above transformations, the Fokker-Planck equation corresponding to
the master equation Eq.\ (\ref{master-anharmonic}) can now be written as
\begin{align}
\frac{\partial }{\partial t}P\left( \hat{\beta}\right) & =\left\{ \frac{%
\partial }{\partial \beta }\left[ \left( \kappa +i\delta \omega +i\lambda
_{11}\right) \beta -2i\lambda _{11}\beta ^{2}\alpha +i\epsilon \right]
-i\lambda _{11}\frac{\partial ^{2}}{\partial \beta ^{2}}\beta ^{2}\right.
\notag \\
& \left. +\frac{\partial }{\partial \alpha }\left[ \left( \kappa -i\delta
\omega -i\lambda _{11}\right) \alpha -2i\lambda _{11}\alpha ^{2}\beta
-i\epsilon \right] -i\lambda _{11}\frac{\partial ^{2}}{\partial \alpha ^{2}}%
\alpha ^{2}+2\kappa N_{1}\frac{\partial ^{2}}{\partial \beta \partial \alpha
}\right\} P\left( \hat{\beta}\right) .  \label{eq:Fokker-Planck}
\end{align}%
The argument of the generalized $P$ function is $\hat{\beta}=\left( \beta
,\alpha \right) ^{T}$. The correspondence principle between operators and $c$%
-numbers is as follows: $\beta \leftrightarrow b$ and $\alpha
\leftrightarrow b^{\dagger }$. However, $\left( \beta ,\alpha \right) $ are
not complex conjugates. Drummond and Gardiner have shown \cite{DG80} that
the Fokker-Planck equation in $\hat{\beta}$ can be transformed to a
stochastic differential equation with positive definite diffusion \footnote{%
Note that their notation is different from ours: their $\beta $ corresponds
to our $\beta $ and their $\beta ^{\dagger }$ to our $\alpha $.}. They found
that the stochastic differential equations in the Ito calculus corresponding
to Eq.\ (\ref{eq:Fokker-Planck}) are
\begin{equation}
\frac{\partial }{\partial t}\left[
\begin{array}{c}
\beta \\
\alpha%
\end{array}%
\right] =\left[
\begin{array}{c}
-i\epsilon -\beta \left( \kappa +i\delta \omega +i\lambda _{11}+2i\lambda
_{11}\beta \alpha \right) \\
i\epsilon -\alpha \left( \kappa -i\delta \omega -i\lambda _{11}-2i\lambda
_{11}\alpha \beta \right)%
\end{array}%
\right] +\left[
\begin{array}{cc}
-2i\lambda _{11}\beta ^{2} & 2\kappa N_{1} \\
2\kappa N_{1} & 2i\lambda _{11}\alpha ^{2}%
\end{array}%
\right] ^{1/2}\left[
\begin{array}{c}
\xi _{1} \\
\xi _{2}%
\end{array}%
\right] ,  \label{diff-eqn0}
\end{equation}%
where $\xi _{1}$ and $\xi _{2}$ are random Gaussian functions, so that $%
\beta $ and $\alpha $ are complex conjugate in the mean \footnote{%
The means of $\beta $ and $\alpha $ are complex conjugates. However,
fluctuation introduces a stochastic component, and so $\beta $ and $\alpha $
deviate from being complex conjugate.}. This stochastic differential
equation is non-linear and not solvable as it is. However, it is reasonable
to use a small noise expansion and linearize the fluctuations about the
steady state of the mean field amplitudes. Thus we write $\beta $ in terms
of the mean amplitude and first order expansion of the fluctuation,
\begin{equation}
\beta \left( t\right) =\beta _{0}+\beta _{1}\left( t\right) ,
\end{equation}%
where $\beta _{0}$ is the steady-state mean amplitude of $\beta $ and is
given by
\begin{equation}
\beta _{0}=\frac{-i\epsilon }{i(\delta \omega +\lambda _{11}+2\lambda
_{11}|\beta _{0}|^{2})+\kappa },  \label{eq:beta0}
\end{equation}%
and $\beta _{1}$ is the zero mean fluctuation amplitude. We have a similar
expression for $\alpha $. Thus $\beta _{0}$\ and $\alpha _{0}$ are complex
conjugate to each other (i.e., $\beta _{0}\alpha _{0}=\left\vert \alpha
_{0}\right\vert ^{2}=\left\vert \beta _{0}\right\vert ^{2}\equiv n_{0}$).
Then to first order in the fluctuations, the fluctuation amplitude vector $%
\hat{\beta}_{1}=(\beta _{1},\alpha _{1})^{T}$ obeys a stochastic
differential equation
\begin{equation}
\frac{\partial }{\partial t}\hat{\beta}_{1}\left( t\right) =-\mathbf{A}\cdot
\hat{\beta}_{1}\left( t\right) +\mathbf{D}^{1/2}\left( \hat{\beta}%
_{0}\right) \hat{\xi}\left( t\right) ,  \label{diff-eqn}
\end{equation}%
where $\hat{\xi}=(\xi _{1},\xi _{2})^{T}$ is the noise vector, $\mathbf{A}$
is the linearized drift matrix and $\mathbf{D}$ is the diffusion matrix
evaluated at $\hat{\beta}=\hat{\beta}_{0}.$ The matrices $\mathbf{A}$ and $%
\mathbf{D}$ are
\begin{equation}
\mathbf{A}=\left[
\begin{array}{cc}
\kappa +i\delta \omega +i\lambda _{11}+4i\lambda _{11}n_{0} & 2i\lambda
_{11}\beta _{0}^{2} \\
-2i\lambda _{11}\alpha _{0}^{2} & \kappa -i\delta \omega -i\lambda
_{11}-4i\lambda _{11}n_{0}%
\end{array}%
\right] ,  \label{eq:A}
\end{equation}%
%
%
%where $n_{0}\equiv \beta _{0}\alpha _{0}$,
and
\begin{equation}
\mathbf{D}=\left[
\begin{array}{cc}
-2i\lambda _{11}\beta _{0}^{2} & 2\kappa N_{1} \\
2\kappa N_{1} & 2i\lambda _{11}\alpha _{0}^{2}%
\end{array}%
\right] .  \label{eq:D}
\end{equation}%
The one-time correlation matrix can be calculated using the method of
Chaturvedi, \textit{et al} \cite{CGMW77,WM,DW80,G85}:%
\begin{align}
\mathbf{C}\left( t,t\right) & =\left[
\begin{array}{cc}
\left\langle \beta _{1}^{2}\right\rangle & \left\langle \beta _{1}\alpha
_{1}\right\rangle \\
\left\langle \alpha _{1}\beta _{1}\right\rangle & \left\langle \alpha
_{1}^{2}\right\rangle%
\end{array}%
\right]  \notag \\
& =\frac{1}{\Lambda ^{2}}\left[
\begin{array}{cc}
-i\lambda _{11}\beta _{0}^{2}\left( \kappa -i\delta \omega -i\lambda
_{11}-4i\lambda _{11}n_{0}\right) \left( 2N_{1}+1\right) & N_{1}\left\vert
\kappa +i\delta \omega +i\lambda _{11}+4i\lambda _{11}n_{0}\right\vert
^{2}+2\lambda _{11}^{2}n_{0}^{2} \\
N_{1}\left\vert \kappa +i\delta \omega +i\lambda _{11}+4i\lambda
_{11}n_{0}\right\vert ^{2}+2\lambda _{11}^{2}n_{0}^{2} & i\lambda
_{11}\alpha _{0}^{2}\left( \kappa +i\delta \omega +i\lambda _{11}+4i\lambda
_{11}n_{0}\right) \left( 2N_{1}+1\right)%
\end{array}%
\right] ,  \label{eq:onetime-corr}
\end{align}%
where%
\begin{eqnarray}
\Lambda ^{2} &=&\kappa ^{2}+\Lambda _{1}^{2},  \label{eq:Lambda} \\
\Lambda _{1}^{2} &=&\left( \delta \omega +\lambda _{11}\right) ^{2}+8\left(
\delta \omega +\lambda _{11}\right) \lambda _{11}n_{0}+12\lambda
_{11}^{2}n_{0}^{2}.  \label{eq:Lambda1}
\end{eqnarray}

We now derive an expression for the two-time steady state correlation matrix%
\begin{equation}
\mathbf{C}\left( t,t^{\prime }\right) =\left[
\begin{array}{cc}
\left\langle \beta _{1}\left( t\right) \beta _{1}\left( t^{\prime }\right)
\right\rangle & \left\langle \beta _{1}\left( t\right) \alpha _{1}\left(
t^{\prime }\right) \right\rangle \\
\left\langle \alpha _{1}\left( t\right) \beta _{1}\left( t^{\prime }\right)
\right\rangle & \left\langle \alpha _{1}\left( t\right) \alpha _{1}\left(
t^{\prime }\right) \right\rangle%
\end{array}%
\right] \;.  \label{2-t-CF}
\end{equation}%
For $t>t^{\prime }$
\begin{equation}
\mathbf{C}\left( t,t^{\prime }\right) =\exp \left( -\mathbf{A}\left(
t-t^{\prime }\right) \right) \mathbf{C}\left( t,t\right) ,
\label{eq:two-time corr}
\end{equation}%
and for $t<t^{\prime }$%
\begin{equation}
\mathbf{C}\left( t,t^{\prime }\right) =\mathbf{C}\left( t,t\right) \exp
\left( -\mathbf{A}^{T}\left( t^{\prime }-t\right) \right) .
\label{eq:two-time corr2}
\end{equation}%
Let us define $\mathbf{M}(t,t^{\prime })\equiv \exp [-\mathbf{A}(t-t^{\prime
})]$. The matrix $\mathbf{M}$ can be calculated as follows. Let the matrix $%
\mathbf{U=}\left( u_{1},u_{2}\right) $ diagonalize $\mathbf{A}$ with
eigenvalues $\lambda _{\pm }$. The eigenvalues for this $2\times 2$ matrix
can be found from the characteristic equation:
\begin{align}
\lambda _{\pm }& =\frac{\mathrm{Tr}\left( \mathbf{A}\right) \pm \sqrt{\left[
\mathrm{Tr}\left( \mathbf{A}\right) \right] ^{2}-4\det \left( \mathbf{A}%
\right) }}{2}  \notag \\
& =\kappa \pm i\Lambda _{1},
\end{align}%
We then obtain the matrix $\mathbf{M}$ as
\begin{eqnarray}
\mathbf{M}(t,t^{\prime }) &=&\boldsymbol{U}\left[
\begin{array}{cc}
\exp \left( -\lambda _{+}\left( t-t^{\prime }\right) \right) & 0 \\
0 & \exp \left( -\lambda _{-}\left( t-t^{\prime }\right) \right)%
\end{array}%
\right] \boldsymbol{U}^{-1}  \notag \\
&=&\frac{1}{2\Lambda _{1}}\left[
\begin{array}{cc}
\left( \Lambda _{1}-c\right) e^{-\lambda _{-}\left( t-t^{\prime }\right)
}+\left( \Lambda _{1}+c\right) e^{-\lambda _{+}\left( t-t^{\prime }\right) }
& 2\lambda _{11}\beta _{0}^{2}\left[ -e^{-\lambda _{-}\left( t-t^{\prime
}\right) }+e^{-\lambda _{+}\left( t-t^{\prime }\right) }\right] \\
2\lambda _{11}\alpha _{0}^{2}\left[ e^{-\lambda _{-}\left( t-t^{\prime
}\right) }-e^{-\lambda _{+}\left( t-t^{\prime }\right) }\right] & \left(
\Lambda _{1}+c\right) e^{-\lambda _{-}\left( t-t^{\prime }\right) }+\left(
\Lambda _{1}-c\right) e^{-\lambda _{+}\left( t-t^{\prime }\right) }%
\end{array}%
\right] ,  \label{eq:M}
\end{eqnarray}%
where $c\equiv 4\lambda _{11}n_{0}+\delta \omega +\lambda _{11}$. The
two-time correlation matrix Eq.\ (\ref{2-t-CF}), then follows directly from
Eqs.\ (\ref{eq:two-time corr}) (\ref{eq:two-time corr2}) and (\ref{eq:M}),
as well as the fact that $\exp [-\mathbf{A}(t-t^{\prime })]=\mathbf{M}%
(t,t^{\prime })$ and $\exp [-\mathbf{A}^{T}(t^{\prime }-t)]=\mathbf{M}%
^{T}(t^{\prime },t)$:

The detailed expressions of the two-time correlation functions are shown in
the Appendix. We note that in the P-representation, the $c$-number time
correlation function corresponds to a normally ordered time correlation
function of the operators; thus the correlations above do not correspond to
all the two-time correlation functions we need to find. For non-normally
ordered time correlation functions, some care needs to be exercised. Using
the procedure described, for example, in Refs.\ \cite{G85, GP00}, we obtain
the following operator to $c$-number correspondence:
\begin{align}
\left\langle b_{1}\left( t\right) b_{1}\left( t^{\prime }\right)
\right\rangle & =\left\langle \beta _{1}\left( t\right) \beta _{1}\left(
t^{\prime }\right) \right\rangle ,  \label{eq:cf1} \\
\left\langle b_{1}\left( t\right) b_{1}^{\dagger }\left( t^{\prime }\right)
\right\rangle & =\left\langle \beta _{1}\left( t\right) \alpha _{1}\left(
t^{\prime }\right) \right\rangle +M_{11}(t,t^{\prime }), \\
\left\langle b_{1}^{\dagger }\left( t\right) b_{1}\left( t^{\prime }\right)
\right\rangle & =\left\langle \alpha _{1}\left( t\right) \beta _{1}\left(
t^{\prime }\right) \right\rangle , \\
\left\langle b_{1}^{\dagger }\left( t\right) b_{1}^{\dagger }\left(
t^{\prime }\right) \right\rangle & =\left\langle \alpha _{1}\left( t\right)
\alpha _{1}\left( t^{\prime }\right) \right\rangle +M_{21}(t,t^{\prime }), \\
\left\langle b_{1}\left( t^{\prime }\right) b_{1}\left( t\right)
\right\rangle & =\left\langle \beta _{1}\left( t\right) \beta _{1}\left(
t^{\prime }\right) \right\rangle +M_{12}(t,t^{\prime }), \\
\left\langle b_{1}\left( t^{\prime }\right) b_{1}^{\dagger }\left( t\right)
\right\rangle & =\left\langle \alpha _{1}\left( t\right) \beta _{1}\left(
t^{\prime }\right) \right\rangle +M_{22}(t,t^{\prime }), \\
\left\langle b_{1}^{\dagger }\left( t^{\prime }\right) b_{1}\left( t\right)
\right\rangle & =\left\langle \beta _{1}\left( t\right) \alpha _{1}\left(
t^{\prime }\right) \right\rangle , \\
\left\langle b_{1}^{\dagger }\left( t^{\prime }\right) b_{1}^{\dagger
}\left( t\right) \right\rangle & =\left\langle \alpha _{1}\left( t\right)
\alpha _{1}\left( t^{\prime }\right) \right\rangle ,  \label{eq:cf8}
\end{align}%
where $M_{ij}(t,t^{\prime })$ are the matrix elements of the matrix $\mathbf{%
M}(t,t^{\prime })$, Eq.\ (\ref{eq:M}).

\section{Master Equation for a reduced density matrix\label{Sec_analysis}}

Having found the one-time and two-time correlation functions, we can now
evaluate Eqs. (\ref{eq:1st order}) and (\ref{2nd order}) and obtain the
master equation for the reduced density matrix of the system oscillator as:
\begin{align}
\frac{d\rho }{dt}& =-i\left( \omega _{0}+\Delta \right) \left[ a^{\dagger
}a,\rho \right] -i\Theta \left[ \left( a^{\dagger }a\right) ^{2},\rho \right]
-\Gamma \left[ a^{\dagger }a,\left[ a^{\dagger }a,\rho \right] \right]
\notag \\
& +\nu \left( N_{0}+1\right) \mathcal{D}\left[ a\right] \rho +\nu N_{0}%
\mathcal{D}\left[ a^{\dagger }\right] \rho  \label{eq:master-final}
\end{align}
where
\begin{align}
\Delta & ={\lambda _{01}}\left[ n_{0}+\frac{1}{\Lambda ^{2}}\left(
N_{1}\left\vert \kappa +i\delta \omega +i\lambda _{11}+4i\lambda
_{11}n_{0}\right\vert ^{2}+2\lambda _{11}^{2}n_{0}^{2}\right) \right] , \\
\Theta & =\lambda _{00}+\frac{\lambda _{01}^{2}n_{0}}{\Lambda ^{2}}\left(
\delta \omega +i\lambda _{11}+2\lambda _{11}n_{0}\right) , \\
\Gamma & =\frac{\lambda _{01}^{2}}{\Lambda ^{4}}\kappa n_{0}(2N_{1}+1)\left[
\left\vert \kappa +i\delta \omega +i\lambda _{11}+4i\lambda
_{11}n_{0}\right\vert ^{2}-4\lambda _{11}n_{0}(\delta \omega +\lambda
_{11}+3\lambda _{11}n_{0})\right]  \notag \\
& =\frac{\lambda _{01}^{2}}{\Lambda ^{4}}\kappa \epsilon ^{2}(2N_{1}+1).
\label{eq:rateG}
\end{align}
We have set $n_{0}=\left\vert \beta _{0}\right\vert ^{2}$, and $\Lambda ^{2}$
is defined in Eqs.\ (\ref{eq:Lambda}) and (\ref{eq:Lambda1}). In obtaining
the last line of Eq.\ (\ref{eq:rateG}), we have used Eq.\ (\ref{eq:beta0}).

In Eq.\ (\ref{eq:master-final}), $\Delta $ in the first term is the resonant
frequency shift due to interactions. The second term is the Kerr non-linear
phase shift, with coefficient $\Theta $ depending on the anharmonicity of
both oscillators $\lambda _{00}$ and $\lambda _{11}$, as well as the
detuning of the ancilla oscillator. The parameter $\Gamma $ is the phase
diffusion coefficient or decoherence rate, associated with back-action due
to an effective measurement of $a^{\dagger }a$. Physically, due to
monitoring, the system would localize or collapse into a phonon number
eigenstate on a time scale of order $\Gamma ^{-1}$. The measurement time
that is needed for the measurement apparatus to distinguish one state from
the next is also proportional to $\Gamma ^{-1}$. The last two terms in Eq.\ (%
\ref{eq:master-final}), can be derived from the thermal coupling to the
system and are responsible for the quantum jumps. In the case when $\nu =0$,
the conditional master equation of Eq.\ (\ref{eq:master-final}) will
describe a QND measurement of the system oscillator phonon number. The time
the system stays in a given phonon number state before making a transition
due to either excitation or relaxation is proportional to $\nu ^{-1}$. To be
a good quantum measurement of a phonon number state, we want the system's
dwelling time to be long compared to the time necessary to determine which
number state the system is in, i.e., $(\Gamma /\nu )\gg 1$.

\subsection{Effects of the anharmonic terms}

From Eq.\ (\ref{eq:master-final}), we notice several important points.
Firstly, in the case of no detuning and no non-linear self-anharmonic terms
(i.e., $\delta \omega =0,$ $\lambda _{00}=\lambda _{11}=0$), we have
\begin{align}
\Delta & =\lambda _{01}\left[ N_{1}+(\epsilon /\kappa )^{2}\right] , \\
\Theta & =0, \\
\Gamma & =\frac{\lambda _{01}^{2}\epsilon ^{2}\left( 2N_{1}+1\right) }{%
\kappa ^{3}}.
\end{align}%
These results agree with the results of a simpler model discussed in Ref.\
\cite{SDC03}, using a slightly different adiabatic elimination approach.

Secondly, the steady state solution Eq.~(\ref{eq:beta0}) of Eq.~(\ref%
{diff-eqn0}) gives
\begin{equation}
\left\vert \epsilon \right\vert ^{2}=n_{0}\left[ \kappa ^{2}+\left( \delta
\omega +\lambda _{11}+2\lambda _{11}n_{0}\right) ^{2}\right] .
\label{eq:n and E relation}
\end{equation}%
Equation (\ref{eq:n and E relation}) has an analogy to a classical
anharmonic oscillator\cite{LL1}. Bistability due to a Kerr nonlinearity is a
well known phenomenon. Classically the oscillator will take one or the other
of the stable solutions. Using Hurwitz stability criterion, to obtain stable
solution for Eqs.~(\ref{diff-eqn})--~(\ref{eq:D}) it is necessary to have
\begin{eqnarray}
\mathrm{Tr}(\mathbf{A}{}) &>&0,  \label{eq:stability} \\
\mathrm{Det}({}\mathbf{A}) &>&0.
\end{eqnarray}%
For the matrix ${}$, Eq.~(\ref{eq:A}), gives $\mathrm{Tr}(\mathbf{A}%
{})=2\kappa >0$ for a dissipative or loss mechanism. Therefore the threshold
points are determined by $\mathrm{Det}(\mathbf{A}{})=\Lambda ^{2}=0$.
However, in the quantum regime at zero temperature, bistability appears only
during transient period and does not exist in the steady state.\cite{WM,
DG80} %\cite{WM, DG80, RonSim}.
We, nevertheless, note that the linear theory that we use to calculate the
steady state correlation functions at finite temperatures would break down
at the instability points.

Secondly, from Eqs.\ (\ref{eq:master-final}) and (\ref{eq:rateG}), we see
that when $\delta \omega =0$, the condition $\kappa \gg \lambda _{11}$ makes
the effect of the non-linear self-anharmonic terms in $\Delta $ and $\Gamma $
very small, which justifies the assumption of neglecting $\lambda _{11}$ in
Ref.\ \cite{SDC03}. However, our calculation allows us to do a quantitative
analysis without making this assumption.

\begin{figure}[tbh]
\begin{center}
\includegraphics[width=4.5in]{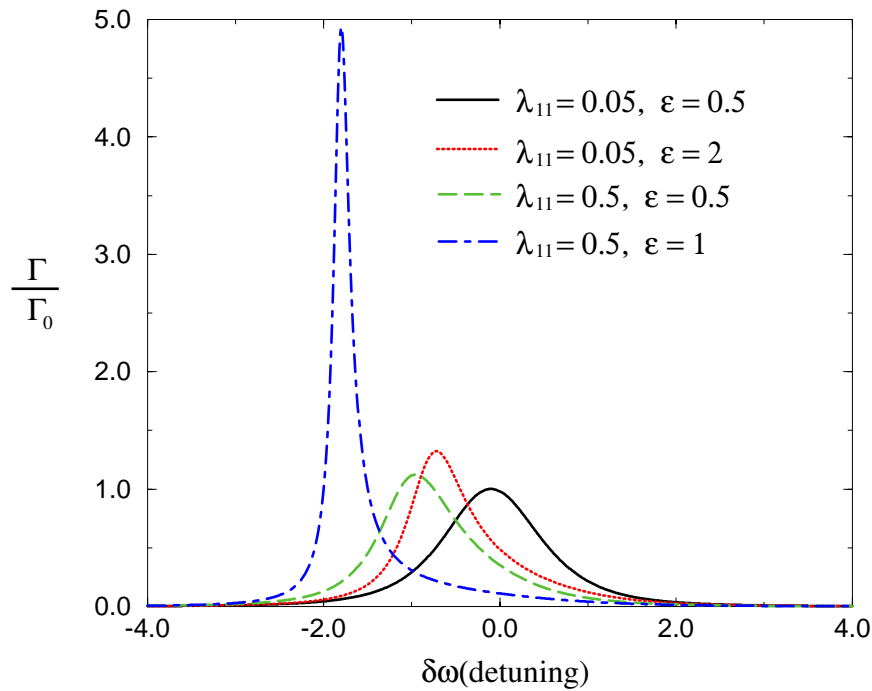}
\end{center}
\caption{The ratio of $(\Gamma/\Gamma_0)$ as a function of detuning at
different values of driving strength and self-anharmonicity (Kerr effect
coupling). The parameters are presented in unit of the damping rate $\protect%
\kappa$.}
\label{QJ_Gamma}
\end{figure}

The value of the phase diffusion coefficient $\Gamma $ (as compared to the
damping rate $\nu $) is important to the phonon number measurement scheme
and to the observation of quantum jumps. To see the effects of
self-anharmonicity, driving and detuning on the phase diffusion coefficient $%
\Gamma $ compared to its value $\Gamma _{0}$ at zero self-anharmonic
coupling and zero detuning ($\lambda _{11}=\epsilon =0$) \cite{SDC03}, we
plot their ratio
\begin{equation}
\frac{\Gamma }{\Gamma _{0}}=\frac{\kappa ^{4}}{\Lambda ^{4}}
\label{eq:QJ_gamma}
\end{equation}%
in Fig.~\ref{QJ_Gamma}. Note that $\Gamma $ diverges at $\Lambda ^{2}=0$,
which are the instability points where the linear theory is not valid. The
parameters (in units of $\kappa $) in Fig.~\ref{QJ_Gamma} are chosen so that
the ancilla oscillator is away from these points. For example, if we were to
increase further the driving strength in the dot-dashed line plot of Fig.~%
\ref{QJ_Gamma}, to $\epsilon =1.2$, say, the ancilla oscillator would then
be in the instability regime. When the nonlinearity $\lambda _{11}$ is
small, the solid line plot in Fig.~\ref{QJ_Gamma} shows the linear resonance
of small driving. The dotted, dashed, and dot-dashed line plots illustrate
that increasing the driving strength and the nonlinearity tends to shift the
resonance frequency, increase the peak value and decrease the width of the
peak of $(\Gamma /\Gamma _{0})$.

Carr and Wybourne have estimated an anharmonic coefficient $\lambda _{ii}$
for a beam with rectangular cross-section\cite{CLW02}:
\begin{equation}
\lambda _{ii}=\frac{\pi ^{4}}{128}\frac{\hbar B}{\rho _{i}^{2}\omega
_{i}^{2}L_{i}^{5}w_{i}t_{i}},  \label{eq:anharmonic-self}
\end{equation}%
where $B$ is the bulk modulus, $\rho _{i}$ is the mass density, $L,w,t$ are
the dimensions of the beam: length, width, thickness, respectively. A simple
estimation of $\kappa $ and $\lambda _{11}$ using realistic values for a
mesoscopic mechanical oscillator reveals that $\lambda _{11}$ is many order
of magnitude smaller than $\kappa $.

\section{Measurement current\label{Sec_current}}

In the measurement scheme, we do not observe the phonon number of the system
oscillator directly. Rather we perform a phase sensitive, `homodyne',
measurement on the quadrature $(b+b^{\dagger })$ of the ancilla oscillator.
It is therefore important to show that an observation of the average current
$\langle I\rangle =\sqrt{2\mu }\langle b+b^{\dagger }\rangle $ indeed
corresponds to a phonon number measurement of the system oscillator. We
anticipate that the average measured current of the ancilla oscillator is
proportional to the average phonon number in the measured system oscillator.
In addition we need to show that the coefficient of proportionality is
related to the localization rate, which determines how long it takes to
distinguish one number state from the next. Thus a strong signal corresponds
to a rapid localization rate. Furthermore we expect that the localization
rate is proportional to the backaction induced phase diffusion coefficient $%
\Gamma$, so that the better the measurement, the larger is the back action
noise.

To demonstrate this, firstly we use the Hamiltonian to obtain the quantum
Langevin equation for the ancilla oscillator operator $b$:
\begin{eqnarray}
\frac{db}{dt} &=&-i\epsilon -i\delta \omega b-i\lambda _{01}a^{\dagger
}ab-i2\lambda _{11}b^{\dagger }bb-\left[ \eta b\left( t\right) -\sqrt{2\eta }%
B_{in}\left( t\right) \right] -\left[ \mu b\left( t\right) -\sqrt{2\mu }%
D_{in}\left( t\right) \right] ,  \label{eq:b} \\
\frac{db^{\dagger }}{dt} &=&i\epsilon +i\delta \omega b^{\dagger }+i\lambda
_{01}a^{\dagger }ab^{\dagger }+i2\lambda _{11}b^{\dagger }b^{\dagger }b-%
\left[ \eta b^{\dagger }\left( t\right) -\sqrt{2\eta }B_{in}^{\dagger
}\left( t\right) \right] -\left[ \mu b^{\dagger }\left( t\right) -\sqrt{2\mu
}D_{in}^{\dagger }\left( t\right) \right] ,  \label{eq:b+}
\end{eqnarray}%
where $B_{in}\left( t\right) $ is the input noise \cite{GP00}. The steady
state [$(db/dt)=0$] average of $\langle b\rangle =\beta _{0}$ for the
ancilla oscillator in isolation (i.e., with $\lambda _{01}=0$) is given by
the same expression as Eq.\ (\ref{eq:beta0}). Linearizing around the steady
state, renaming the operator describing the quantum fluctuation as $b_{1}(t)$%
, and assuming that $a^{\dagger }a$ do not to change appreciably over the
typical time scale of the ancilla oscillator, we obtain
\begin{eqnarray}
\frac{db_{1}}{dt} &=&-i(\delta \omega +\lambda _{11})b_{1}+2\lambda
_{11}(2n_{0}b_{1}+\beta _{0}^{2}b_{1}^{\dagger })+\lambda _{01}\beta
_{0}a^{\dagger }a-\kappa b_{1}+\sqrt{2\eta }B_{in}+\sqrt{2\mu }D_{in} \\
\frac{db_{1}^{\dagger }}{dt} &=&i(\delta \omega +\lambda
_{11})b_{1}^{\dagger }+2\lambda _{11}(2n_{0}b_{1}^{\dagger }+(\alpha
_{0}^{2}b_{1}^{\dagger })+\lambda _{01}\alpha _{0}a^{\dagger }a-\kappa
b_{1}^{\dagger }+\sqrt{2\eta }B_{in}^{\dagger }+\sqrt{2\mu }D_{in}^{\dagger
}.
\end{eqnarray}%
or equivalently,
\begin{equation}
\frac{d}{dt}\left(
\begin{array}{c}
b_{1} \\
b_{1}^{\dagger }%
\end{array}%
\right) =\mathbf{A}\left(
\begin{array}{c}
b_{1} \\
b_{1}^{\dagger }%
\end{array}%
\right) +\left(
\begin{array}{c}
\lambda _{01}\beta _{0}a^{\dagger }a+\sqrt{2\eta }B_{in}+\sqrt{2\mu }D_{in}
\\
\lambda _{01}\alpha _{0}a^{\dagger }a+\sqrt{2\eta }B_{in}^{\dagger }+\sqrt{%
2\mu }D_{in}^{\dagger }%
\end{array}%
\right) ,  \label{eq:b1}
\end{equation}%
where $\mathbf{A}$ is defined in Eq.\ (\ref{eq:A}). To calculate $\langle
b+b^{\dagger }\rangle =\beta _{0}+\alpha _{0}+\langle b_{1}+b_{1}^{\dagger
}\rangle $ in the steady state, we setting $(db_{1}/dt)=0=(db_{1}^{\dagger
}/dt)$ in Eq.\ (\ref{eq:b1}), to obtain
\begin{equation}
\left(
\begin{array}{c}
\langle b_{1}\rangle \\
\langle b_{1}^{\dagger }\rangle%
\end{array}%
\right) =\mathbf{A}^{-1}\left(
\begin{array}{c}
\lambda _{01}\beta _{0}a^{\dagger }a \\
\lambda _{01}\alpha _{0}a^{\dagger }a%
\end{array}%
\right) .
\end{equation}%
Then after a simple calculation, we obtain the measured mean signal
\begin{eqnarray}
\sqrt{2\mu }\,\langle b_{1}+b_{1}^{\dagger }\rangle &=&-i\sqrt{2\mu }\,\frac{%
\lambda _{01}}{\Lambda ^{2}}\left\{ \left[ \left( \kappa -i(\delta \omega
+\lambda _{11}+4\lambda _{11}|\beta _{0}|^{2})+2i\lambda _{11}(\alpha
_{0})^{2}\right) \beta _{0}\right] -h.c.\right\} \langle a^{\dagger }a\rangle
\notag \\
&=&-i\sqrt{2\mu }\,\frac{\lambda _{01}}{\Lambda ^{2}}\left[ \kappa (\beta
_{0}-\alpha _{0})-i(\delta \omega +\lambda _{11}+2\lambda _{11}|\beta
_{0}|^{2})(\beta _{0}+\alpha _{0})\right] \langle a^{\dagger }a\rangle .
\label{eq:avg_b1}
\end{eqnarray}%
Using Eq.\ (\ref{eq:beta0}), we can simplify Eq.\ (\ref{eq:avg_b1}) further
and obtain
\begin{equation}
\sqrt{2\mu }\,\langle b_{1}+b_{1}^{\dagger }\rangle =-\sqrt{2\mu }\,\frac{%
2\epsilon \,\lambda _{01}}{\Lambda ^{2}}\langle a^{\dagger }a\rangle .
\label{eq:current}
\end{equation}%
We note that the coefficient on the right hand side of Eq.(\ref{eq:current})
is proportional to $\sqrt{\Gamma }$, with a proportionality factor given by $%
-\sqrt{8\mu /\kappa (2N_{1}+1)}$. As the actual readout current is simply
proportional to the average position of the ancilla oscillator \cite{R00},
Eq.(\ref{eq:current}) gives the expected proportionality between the average
measured current and the average phonon number of the system oscillator.

In a typical experimental run, the measured current will contain a noise
component made up of thermo-electrical noise in the transducer circuit as
well as intrinsic quantum noise that arises directly from the back action
noise when we measure phonon number. In order for the measurement to be
quantum limited, we need to ensure that the dominant source of noise is back
action noise. Recently, considerable progress towards this limit has been
made in a nanoelectromechanical system \cite{Schwab04}

\section{Conclusions}

We have investigated a scheme for the QND measurement of phonon number (cf\
\cite{SDC03}) using two anharmonically coupled modes of oscillation of
mesoscopic elastic bridges. We have included the self-anharmonic terms
neglected in the previous analysis \cite{SDC03}, and analyzed the effect of
higher order anharmonic terms in the approximation that the ancilla
oscillator is heavily damped. We have shown that in the presence of
self-anharmonic term, $x_{1}^{4}$, of the ancilla oscillator, the effect of
increasing driving strength and self-nonlinearity tends to shift the
resonance frequency, increase the peak value and decrease the width of the
response of the peak of $(\Gamma /\Gamma _{0})$ as shown in Fig.~\ref%
{QJ_Gamma}. If the damping of the ancilla oscillator is much larger than the
effect of the self-anharmonic term, the overall effect of self-anharmonic
term on the phonon number measurement is small for small detuning,
justifying the assumption of neglecting the self-anharmonic term at zero
detuning in Ref.~\cite{SDC03}. Our calculation, however, allows one to do a
quantitative analysis at finite detuning and without making this assumption.

The key idea of the measurement scheme is that, from the point of view of
the ancilla oscillator, the interaction with the system oscillator
constitutes a shift in resonance frequency that is proportional to the
time-averaged phonon number or energy excitation of the system oscillator.
This frequency shift may be detected through a phase sensitive readout of
the position of the driven readout oscillator. In a magnetic field, a wire
patterned on the moving readout oscillator will result in an induced current
which can be directly monitored by electrical means \cite{YGPB94}. The
current gives direct access to the position of the ancilla oscillator and,
through the mechanism described in this paper, to the phonon number of the
measured system oscillator, even in the presence of the self-anharmonic
terms. We have shown that this scheme realizes an ideal QND measurement of
phonon number in the limit that the back action induced phase diffusion rate
is much larger than the rate at which transitions occur between phonon
number states, $(\Gamma /\nu )\rightarrow \infty $. When the ratio $(\Gamma
/\nu )$ is finite and large, it is then possible to observe, in the readout
current, quantum jumps between Fock (number states) in a mesoscopic
mechanical oscillator, as the mechanical oscillator exchanges quanta with
the environment.

We briefly discuss below some possible realistic values for $\Gamma $ and $%
\nu$. The value of $\Gamma _{0}$ depends on external driving, as well as
materials and dimensions of the mechanical beams (oscillators). Here we
quote the example in Ref.~\cite{SDC03} using two GaAs mechanical oscillators
with resonance frequencies $\omega _{0}=2.3$ GHz, $\omega _{1}=0.36$ GHz,
and Q-factors $Q_{0}=10000$, $Q_{1}=1000$. The dimensions of the system
oscillator are $0.6$ $\mu $m $\times $ $0.04$ $\mu $m $\times $ $0.07$ $\mu $%
m and those of the ancilla oscillator are $0.6$ $\mu $m $\times $ $0.04$ $%
\mu $m $\times $ $0.01$ $\mu $m. With the magnetic field $10$ Tesla and the
driving current $1$ $\mu $A, $\Gamma _{0}$ and $\nu $ will be $\Gamma
_{0}\approx 1.5\times 10^{4}$/s and $\nu \approx 1.2\times 10^{6}$/s, or $%
\Gamma _{0}/\nu =0.013$. A clear observation of quantum jumps requires $%
\Gamma _{0}/\nu \gg 1$, so that the present example is two orders of
magnitude below the desired parameter regime. To increase the ratio of $%
\Gamma$ to $\nu$ we can improve on some of the parameters. One way is to
increase the Q-factor of the system oscillator. Another way is to use lower
density material such as carbon nanotubes as well as to decrease the
thickness of the oscillator. These improvements are feasible with current
fabrication technology. In addition, it is also possible to engineer the
nonlinear coupling between the oscillators \cite{SP:Y}. Furthermore,
different driving and detection schemes other than magnetomotive detection
can be considered to increase the driving strength. Given the steady
improvement in the fabrication technology and experimental techniques, we
believe that observing quantum jumps between phonon number states in a
mesoscopic oscillator will be possible in the near future.

\begin{acknowledgments}
DHS is grateful to the SRC for Quantum Computer Technology at the University
of Queensland for their hospitality during her extensive stay and thanks
Michael Cross for useful discussions. DHS's work is supported by DARPA
DSO/MOSAIC through grant N00014-02-1-0602 and by the NSF through a grant for
the Institute for Theoretical Atomic, Molecular and Optical Physics at
Harvard University and Smithsonian Astrophysical Observatory. HSG would like
to acknowledge financial support from Hewlett-Packard.
\end{acknowledgments}

\appendix

\section{Expressions for the two-time correlation functions}

The two-time correlation functions in the main text for $C\left( t,t^{\prime
}\right) $, where $t>t^{\prime }$, are
\begin{align}
\left\langle \beta _{1}\left( t\right) \beta _{1}\left( t^{\prime }\right)
\right\rangle & =\frac{1}{2\Lambda _{1}}\left\{ \left( \Lambda _{1}+c\right)
\exp \left[ -\lambda _{+}\left( t-t^{\prime }\right) \right] +\left( \Lambda
_{1}-c\right) \exp \left[ -\lambda _{-}\left( t-t^{\prime }\right) \right]
\right\} \left\langle \beta _{1}^{2}\right\rangle  \notag \\
& +\frac{\lambda _{11}\beta _{0}^{2}}{\Lambda _{1}}\left\{ \exp \left[
-\lambda _{+}\left( t-t^{\prime }\right) \right] -\exp \left[ -\lambda
_{-}\left( t-t^{\prime }\right) \right] \right\} \left\langle \beta
_{1}\alpha _{1}\right\rangle ,
\end{align}%
\begin{align}
\left\langle \alpha _{1}\left( t\right) \beta _{1}\left( t^{\prime }\right)
\right\rangle & =\frac{1}{2\Lambda _{1}}\left\{ \left( \Lambda _{1}+c\right)
\exp \left[ -\lambda _{+}\left( t-t^{\prime }\right) \right] +\left( \Lambda
_{1}-c\right) \exp \left[ -\lambda _{-}\left( t-t^{\prime }\right) \right]
\right\} \left\langle \alpha _{1}\beta _{1}\right\rangle  \notag \\
& +\frac{\lambda _{11}\beta _{0}^{2}}{\Lambda _{1}}\left\{ \exp \left[
-\lambda _{+}\left( t-t^{\prime }\right) \right] -\exp \left[ -\lambda
_{-}\left( t-t^{\prime }\right) \right] \right\} \left\langle \alpha
_{1}^{2}\right\rangle ,  \label{eq:betatalphatp}
\end{align}%
\begin{align}
\left\langle \beta _{1}\left( t\right) \alpha _{1}\left( t^{\prime }\right)
\right\rangle & =\frac{-\lambda _{11}\alpha _{0}^{2}}{\Lambda _{1}}\left\{
\exp \left[ -\lambda _{+}\left( t-t^{\prime }\right) \right] -\exp \left[
-\lambda _{-}\left( t-t^{\prime }\right) \right] \right\} \left\langle \beta
_{1}^{2}\right\rangle  \notag \\
& +\frac{1}{2\Lambda _{1}}\left\{ \left( \Lambda _{1}-c\right) \exp \left[
-\lambda _{+}\left( t-t^{\prime }\right) \right] +\left( \Lambda
_{1}+c\right) \exp \left[ -\lambda _{-}\left( t-t^{\prime }\right) \right]
\right\} \left\langle \beta _{1}\alpha _{1}\right\rangle ,
\end{align}%
\begin{align}
\left\langle \alpha _{1}\left( t\right) \alpha _{1}\left( t^{\prime }\right)
\right\rangle & =\frac{-\lambda _{11}\alpha _{0}^{2}}{\Lambda _{1}}\left\{
\exp \left[ -\lambda _{+}\left( t-t^{\prime }\right) \right] -\exp \left[
-\lambda _{-}\left( t-t^{\prime }\right) \right] \right\} \left\langle
\alpha _{1}\beta _{1}\right\rangle  \notag \\
& +\frac{1}{2\Lambda _{1}}\left\{ \left( \Lambda _{1}-c\right) \exp \left[
-\lambda _{+}\left( t-t^{\prime }\right) \right] +\left( \Lambda
_{1}+c\right) \exp \left[ -\lambda _{-}\left( t-t^{\prime }\right) \right]
\right\} \left\langle \alpha _{1}^{2}\right\rangle ,
\end{align}%
where $c=4\lambda _{11}n_{0}+\delta \omega +\lambda _{11}$ and $\Lambda
_{1}^{2}=\left( \delta \omega +\lambda _{11}\right) ^{2}+8\left( \delta
\omega +\lambda _{11}\right) \lambda _{11}n_{0}+12\lambda _{11}^{2}n_{0}^{2}$
as in the main text. These equations give the c-number two-time correlation
functions we need to obtain the operator two-time correlation functions in
Eq.~(\ref{eq:cf1})--(\ref{eq:cf8}).

%%%%%%%%%%%%%%%%%%%%%%%%%%%%%%%%%%%%%%%%%%%%%%%%%%%%%%%%%%%%%%%%%%%%%%%%%
%%%%%%%%%%%%%%%%%%%%%%%%%%%%%%%%%%%%%%%%%%%%%%%%%%%%%%%%%%%%%%%%%%%%%%%%%

\end{document}